\pdfoutput=1
\documentclass[11pt,a4paper]{article}
\usepackage[hyperref]{naaclhlt2019  }
\usepackage{times}
\usepackage{latexsym}
\usepackage{amsmath}
\usepackage{amssymb}
\usepackage{fancyhdr}
\usepackage{hyperref}
\usepackage{graphicx}
\usepackage{natbib}
\usepackage{hhline}
\usepackage{booktabs, multirow}
\usepackage{float}
\usepackage{url}
\usepackage[show]{chato-notes}

\usepackage{color}

\aclfinalcopy 

\title{Team QCRI-MIT at SemEval-2019 Task 4:\\ Propaganda Analysis Meets Hyperpartisan News Detection}

\author{%
	Abdelrhman Saleh$^1$, Ramy Baly$^2$, Alberto Barr\'on-Cede\~no$^3$,\\
	\bf Giovanni Da San Martino$^3$, Mitra Mohtarami$^2$, Preslav Nakov$^3$, James Glass$^2$\\
	$^1$Harvard University, MA, USA\\
    $^2$MIT Computer Science and Artificial Intelligence Laboratory, MA, USA\\
    $^3$Qatar Computing Research Institute, HBKU, Qatar\\
    {\tt abdelrhman\_saleh@college.harvard.edu},\\
    {\tt \{baly, mitram, glass\}@mit.edu}\\
    {\tt \{albarron, gmartino, pnakov\}@hbku.edu.qa}
 }

\date{}

\begin{document}
\maketitle
\begin{abstract}
  In this paper, we describe our submission to SemEval-2019 Task 4 on Hyperpartisan News Detection.
  Our system relies on a variety of engineered features originally used to detect propaganda.
  This is based on the assumption that biased messages are propagandistic in the sense that they promote a particular political cause or viewpoint.
  We trained a logistic regression model with features ranging from simple bag-of-words to vocabulary richness and text readability features.
  Our system achieved 72.9\% accuracy on the test data that is annotated manually and 60.8\% on the test data that is annotated with distant supervision.
  Additional experiments showed that significant performance improvements can be achieved with better feature pre-processing.\footnote{Our system is available at \url{https://github.com/AbdulSaleh/QCRI-MIT-SemEval2019-Task4}}
\end{abstract}

\section{Introduction}

The rise of social media has enabled people to easily share information with a large audience without regulations or quality control.
This has allowed malicious users to spread disinformation and misinformation (a.k.a. ``fake news'') at an unprecedented rate.
Fake news is typically characterized as being hyperpartisan (one-sided), emotional and riddled with lies~\cite{potthast2017stylometric}.
The SemEval-2019 Task~4 on Hyperpartisan News Detection~\cite{johannes} focused on the challenge of automatically identifying whether a text is hyperpartisan or not.
While hyperpartisanship is defined as ``exhibiting one or more of blind, prejudiced, or unreasoning allegiance to one party, faction, cause, or person'', we model this task as a binary document classification problem.

Scholars have argued that all biased messages can be considered propagandistic, regardless of whether the bias was intentional or not~\cite[p. XV]{Ellul:65}.
As a result, we approached the task departing from an existing model for propaganda identification \cite{AAAI2019:proppy}.
Our hypothesis is that as propaganda is inherent in hyperpartisanship -- the two problems are two sides of the same coin, and solving one of them would help solve the other.
Our system consists of a logistic regression model that is trained with a variety of engineered features that range from word and character TFiDF $n$-grams and lexicon-based features to more sophisticated features that represent different aspects of the article's text such as the richness of its vocabulary and the complexity of its language.

Our official submission achieved an accuracy of 72.9\% (while the winning system achieved 82.2\%).
This was achieved using word and character $n$-grams.
Additional, post-submission experiments show that further performance improvements can be achieved by careful pre-processing of the engineered features.

\section{Related Work}

The analysis of bias and disinformation has attracted significant attention, especially after the 2016 US presidential election~\cite{brill2001online,finberg2002digital,Castillo:2011:ICT:1963405.1963500,D18-1389,Kulkarni:2018:EMNLP,InternetResearchJournal:2018}.
Most of the proposed approaches have focused on predicting credibility, bias or stance.
\citet{Popat:2017:TLE:3041021.3055133} assessed the credibility of claims based on the occurrence of assertive and factive verbs, hedges, implicative words, report verbs and discourse markers, which were extracted using manually crafted gazetteers (referred to as stylistic features).

Stance detection was considered as an intermediate step for detecting fake claims, where the veracity of a claim is checked by aggregating the stances of retrieved relevant articles~\cite{baly-EtAl:2018:N18-2}.
Several stance detection models have been proposed as part of the Fake News Challenge (FNC)\footnote{\url{http://www.fakenewschallenge.org}} including deep convolutional neural networks~\cite{talos:2017:fnc}, multi-layer perceptrons~\cite{hanselowski2018retrospective}, and end-to-end memory networks~\cite{NAACL2018:stance}

The stylometric analysis model of~\citet{Koppel:07} was used by~\citet{DBLP:journals/corr/PotthastKRBS17} when looking for hyperpartisanship.
They used articles from nine news sources whose factuality has been manually verified by professional journalists.
Writing style and complexity was also considered by~\citet{Horne:17} to differentiate real news from fake news and satire.
They used features such as the number of occurrences of different part-of-speech tags, swearing and slang words, stop words, punctuation, and negation as stylistic markers.
They also used a number of readability measures.
\citet{rashkin-EtAl:2017:EMNLP2017} focused on a multi-class setting: real news, satire, hoax, or propaganda. Their supervised model relied on word $n$-grams.

Similarly to~\citet{DBLP:journals/corr/PotthastKRBS17}, we believe that there is an inherent style in propaganda, regardless of the source publishing it.
Many stylistic features were proposed for authorship identification, i.e., the task of predicting whether a piece of text has been written by a particular author.
One of the most successful representations for such a task are character-level $n$-grams~\cite{Stamatatos:2009}, and they turn out to represent some of our most important stylistic features.


More details about research on fact-checking and the spread of fake news online can be found in~\cite{lazer2018science,Vosoughi1146,thorne-vlachos:2018:C18-1}.

\section{System Description}
\label{sec:System}

We developed our system for detecting hyper-partisanship in news articles by training a logistic regression classifier using a set of engineered features that included the following: character and word $n$-grams, lexicon-based indicators, and readability and vocabulary richness measures.
Below, we describe these features in detail.

\paragraph{Character $3$-grams.}

\citet{Stamatatos:2009} argued that, for tasks where the topic is irrelevant, character-level representations are more sensitive than token-level ones.
We hypothesize that this applies to hyperpartisan news detection, since articles on both sides of the political spectrum may be discussing the same topics.
\citet{Stamatatos:2009} found that ``the most frequent character $n$-grams are the most important features for stylistic purposes''.
These features capture different style markers, such as prefixes, suffixes and punctuation marks.
Following the analysis in~\citet{Barron:19}, we include TFiDF-weighted character 3-grams in our feature set.

\paragraph{Word $n$-grams}
Bag-of-words (BoW) features are widely used for text classification.
We extracted the $k$ most frequent $[1,2]$-grams, and we represented them using their TFiDF scores.
We ignored $n$-grams that appeared in more than 90\% of the documents, most of which contained stopwords and were irrelevant with respect to hyperpartisanship.
Furthermore, we incorporated Naive Bayes by weighing the $n$-grams based on their importance for classification, as proposed by~\citet{wang2012baselines}.
We define $\boldsymbol{x}_i\in\mathbb{R}^{|V|}$ as a row vector in the TFiDF feature matrix, representing the $i^{th}$ training sample with a target label $y_i\in\{0,1\}$, where $V$ is the vocabulary size.
We also define vectors $\boldsymbol{p}=\alpha+\sum_{i:y_i=1} \boldsymbol{x}_i$ and $\boldsymbol{q}=\alpha+\sum_{i:y_i=0} \boldsymbol{x}_i$, and we set the smoothing parameter $\alpha$ to 1.
Finally, we calculate the vector:

\begin{equation}
    \boldsymbol{r} = \log\left(\frac{\boldsymbol{p}/\parallel\boldsymbol{p}\parallel}{\boldsymbol{q}/\parallel\boldsymbol{q}\parallel}\right)
\end{equation}
which is used to scale the TFiDF features to create the NB-TFiDF features as follows:
\begin{equation}
    \boldsymbol{x}'_i = \boldsymbol{r} \circ \boldsymbol{x}_i, \,\quad \forall i
\end{equation}

\paragraph{Bias Analysis}
We analyze the bias in the language used in the documents by
(\emph{i})~creating bias lexicons that contain {\it left} and {\it right} bias cues, and
(\emph{ii})~using these lexicons to compute two scores for each document, indicating the intensity of bias towards each ideology.
To generate the list of cues that signal biased language, we use Semantic Orientation (SO)~\cite{Turney:2002} to identify the words that are strongly associated with each of the left and right documents in the training dataset.
Those SO values can be either positive or negative, indicating association with right or left biases, respectively.
Then, we select words whose absolute SO value is $\geq0.4$ to create two bias lexicons: $BL_{left}$ and $BL_{right}$.
Finally, we use these lexicons to compute two bias scores per document according to Equation~\eqref{Ling-equation}, where for each document $D_j$, the frequency of cues in the lexicon $BL_i$ that are present in $D_j$ is normalized by the total number of words in $D_j$:

\begin{equation}
\label{Ling-equation}
bias_i(D_j) = \dfrac{\displaystyle\sum\limits_{cue \in BL_i} {count(cue, D_j)}}{\displaystyle\sum\limits_{w_k \in D_j} {count(w_k, D_j)}}
\end{equation}

\paragraph{Lexicon-based Features.}
\citet{rashkin-EtAl:2017:EMNLP2017} studied the occurrence of specific types of words in different kinds of articles, and showed that words from certain lexicons (e.g., negation and swear words) appear more frequently in propaganda, satire, and hoax articles than in trustworthy articles.
We capture this by extracting features that reflect the frequency of words from particular lexicons.
We use 18 lexicons from the Wiktionary, Linguistic Inquiry and Word Count (LIWC)~\cite{pennebaker2001linguistic}, Wilson's subjectives~\cite{wilson2005recognizing}, Hyland's hedges~\cite{doi:10.1002/9781118611463.wbielsi003}, and Hooper's assertives~\cite{hooper1974assertive}.
For each lexicon, we count the total number of words in the article that appear in the lexicon.
This resulted in 18 features, one for each lexicon.

\paragraph{Vocabulary Richness}
\citet{DBLP:journals/corr/PotthastKRBS17} showed that hyperpartisan outlets tend to use a writing style that is different from mainstream outlets.
Different topic-independent features have been proposed to characterize the vocabulary richness, style and complexity of a text.
For this task, we used the following vocabulary richness features:
(\emph{i})~type--token ratio (\emph{TTR}): the ratio of types to tokens in a text,
(\emph{ii})~\emph{Hapax Legomena}: number of types appearing once in a text,
(\emph{iii})~\emph{Hapax Dislegomena}: number of types appearing twice in a text,
(\emph{iv})~\emph{Honore's R}: A combination of types, tokens and hapax legomena~\cite{Honore:79}:

\begin{equation}
\text{Honore's R} = \frac{100\times\log(|\text{tokens}|)}{1 - |\text{Legomena}| / |\text{types}|}\label{eq:R} \enspace
\end{equation}

and (\emph{v})~\emph{Yule's characteristic K}: The chance of a word occurring in a text following a Poisson distribution~\cite{Yule:44}:

\begin{equation}
    \text{Yule's K} = 10^4 \cdot \frac{\displaystyle\sum_i i^2|\text{types}_i|-|\text{tokens}|}{|\text{tokens}|^2}\label{eq:Y},
\end{equation}
where tokens refer to all words in a text (including repetitions), types refer to distinct words, $i$ are the tokens' frequency ranks (1 being the least frequent), and types$_i$ are the number of tokens with the $i^{th}$ frequency.

\paragraph{Readability}
We also used the following readability features that were originally designed to estimate the level of text complexity:
1) \emph{Flesch--Kincaid grade level}: represents the US grade level necessary to understand a text~\cite{Kincaid:75},
2) \emph{Flesch reading ease}: is a score for measuring how difficult a text is to read~\cite{Kincaid:75}, and
3) \emph{Gunning fog index}: estimates the years of formal education necessary to understand a text~\cite{Gunning:68}.

\newcommand{\rightcurved}{$\rotatebox[origin=c]{180}{$\Lsh$}$}

\begin{table*}[t]
\centering
\begin{tabular}{cp{3.75cm}|ccccccccc} %
\toprule
 & \multirow{1}{*}{\parbox[c][1.1cm]{0.4\linewidth}{\bf Features}} &
 \multicolumn{4}{c}{Labeled by-{\bf Article}} &&
 \multicolumn{4}{c}{Labeled by-{\bf Publisher}} \\
 \cmidrule{3-6} \cmidrule{8-11}
 & & Accuracy & Prec. & Rec. & F1  &&  Accuracy & Prec. & Rec. & F1 \\
 \midrule
 1 & BoW (TFiDF)                                     & 67.8 & 53.8 & 89.1 & 67.1   &&   56.7  & 55.1  & 72.5  & 62.6 \\
 2 & BoW (NB-TFiDF)                                  & 69.6 & 56.1 & 80.7 & 66.2   &&   57.1  & 56.4  & 61.9  & 59.0 \\
 3 & \rightcurved + Char trigrams                    & 74.0 & 62.5 & 73.5 & 67.6   &&   54.8  & 54.3  & 60.8  & 57.4 \\
 4 & \hspace{0.2cm}\rightcurved + Bias               & 75.2 & 67.7 & 62.6 & 65.1   &&   54.5  & 55.0  & 50.4  & 52.6 \\
 5 & \hspace{0.4cm}\rightcurved + Lexical            & 75.2 & 67.0 & 64.7 & 65.8   &&   52.3  & 52.3  & 51.5  & 51.9 \\
 6 & \hspace{0.6cm}\rightcurved + Vocab. Richness    & 75.8 & 67.1 & 67.6 & 67.4   &&   50.9  & 50.8  & 52.5  & 51.7 \\
 7 & \hspace{0.8cm}\rightcurved + Readability        & 76.0 & 66.4 & 70.6 & 68.4   &&   51.6  & 51.5  & 53.9  & 52.7 \\
 \bottomrule
 \end{tabular}
\caption{\label{tbl:results}An incremental analysis showing the performance of different feature combinations, evaluated on the validation sets labeled by {\it article} and by {\it publisher}.}
\end{table*}

\section{Experiments and Results}

\subsection{Dataset}
We trained our models on the Hyperpartisan News Dataset from SemEval-2019, Task 4~\cite{johannes}, which is split by the task organizers into:
1) {\it Labeled by-Publisher:} contains 750K articles labeled via distant supervision, i.e. using labels of their publisher\footnote{Publishers labels are identified by \href{http://buzzfeed.com}{BuzzFeed} journalists or by the \href{https://mediabiasfactcheck.com}{Media Bias/Fact Check} project}.
Labels are evenly distributed across the ``hyperpartisan'' and ``not-hyperpartisan'' classes.
This set is further split into 600K for training and 150K for validation.
2) {\it Labeled by-Article:} This set contains 645 articles labeled through crowd-sourcing (37\% are hyperpartisan and 63\% are not).
Only articles with a consensus among annotators were included.

\subsection{Experimental Setting}
We train a logistic regression (LR) model with a Stochastic Average Gradient solver~\cite{schmidt2017minimizing} due to the large size of the dataset.
In order to reduce overfitting we use $\mathbb{L}_2$ regularization (with $C=1$ as the regularization parameter).
Feature normalization was needed since the different features represent different aspects of text, hence have very different scales.
We tried to normalize each feature set by subtracting the mean and scaling it to unit variance.
However, we found that multiplying the features by constant scaling factors resulted in better performance.
The scaling factor for each family of features was a hyperparameter that was tuned during the fine-tuning experiments.

We trained the classifier using the 600K training examples annotated {\it by-Publisher}, then used the remaining 150K examples for evaluation.
We fine-tuned the hyperparameters on the 645 {\it by-Article} examples.
The hyper-parameters include $k\in[50, 200, 700]^{\times10^3}$ as the most frequent word $n$-grams and the scaling parameters of the different features except for the $n$-grams.
Best fine-tuning results suggested using the 200K most-frequent word $[1,2]$-grams.
We assessed the different feature sets, described in Section~\ref{sec:System}, by incrementally adding each set, one at a time, to the mix of all features.

\subsection{Results}
Table~\ref{tbl:results} illustrates the results obtained on both the {\it by-Article} set (which we used to fine-tune the model's hyperparameters) and the {\it by-Publisher} set (which we used for evaluation).
Our results suggest that scaling the TFiDF values through Naive Bayes is better than using raw TFiDF scores.
Hence, these features were used for all subsequent experiments.
It can also be observed that adding each group of features introduces a consistent improvement in accuracy on the {\it by-Article} data.
However, we observed an opposite behaviour on the {\it by-Publisher} data.
We believe this is due to the significant amount of noisy labels introduced by the distant supervision labeling strategy.
Therefore, we based our decisions on the results obtained on the {\it by-Article} data since its labels are more accurate.

The normalization strategy, i.e., scaling the features using calibrated scaling parameters, introduced significant performance improvements.
Unfortunately, we were not able to perform these calibration experiments by the competition's deadline, hence we submitted the system that was available at that time, which is based on the BoW (NB-TFiDF) and character 3-gram features, as shown in row 3 in Table~\ref{tbl:results}.
Our system achieved a 72.9\% accuracy on the test {\it by-Article} data, ranking 20\textsuperscript{th}/42.
It also achieved 60.8\% accuracy on the test {\it by-Publisher} data, ranking 15\textsuperscript{th}/42.
All subsequent, and superior, results (rows 4--7) were obtained after the deadline.

\section{Conclusion}

In this paper, we present our submission to SemEval-2019 Task 4 on Hyperpartisan News Detection.
We trained a logistic regression model with a feature set that included word and character $n$-grams, represented with TFiDF.
This system achieved a 72.9\% and 60.8\% accuracy on the test data that is labeled {\it by-Article} and {\it by-Publisher}, respectively.

We also evaluated additional features that represent different aspects of the article's text such as its vocabulary richness, the kind of language it uses according to different lexicons, and its level of complexity.
Initial experiments showed that these features hurt the model.
However, with proper pre-processing and scaling we were able to achieve significant performance improvements of up to 2\% in absolute accuracy.
These results were obtained after the competition's deadline, hence were not considered as part of our submission.

\section{Acknowledgment}

This research was carried out in collaboration be- tween the MIT Computer Science and Artificial Intelligence Laboratory (CSAIL) and the HBKU Qatar Computing Research Institute (QCRI).

\bibliography{references}
\bibliographystyle{acl_natbib}

\end{document}